\documentstyle[aps,twocolumn,prl,epsf]{revtex}

\begin{document}

\draft

\title{An Explicit Space-time Adaptive Method for 
Simulating Complex Cardiac Dynamics}

\author{
Elizabeth~M.\ Cherry${}^\ast$,
Henry~S.\ Greenside${}^\dagger$,
and
Craig~S.\ Henriquez${}^\ddagger$
}

\address{${}^\ast$Department of Computer Science, Duke
U., P.~O.~Box 90129, Durham, NC 27708-0129 }
\address{${}^\dagger$Department of Physics, Duke U.,
Durham, NC 27708-0305}
\address{${}^\ddagger$Department of Biomedical Engineering,
Duke U., Durham, NC 27708-0281}

\date{\today}

\maketitle

\begin{abstract}
For plane-wave and many-spiral states of the experimentally
based Luo-Rudy~1 model of heart tissue in large (8~cm square)
domains, we show that an explicit space-time-adaptive
time-integration algorithm can achieve an order of magnitude
reduction in computational effort and memory---but without a
reduction in accuracy---when compared to an algorithm using
a uniform space-time mesh at the finest resolution. Our
results indicate that such an explicit algorithm can 
be extended straightforwardly to simulate quantitatively 
large-scale three-dimensional electrical dynamics 
over the whole human heart.
\end{abstract}
\pacs{
05.45.Pq, % Numerical simulations of chaotic models
87.15.Aa, % Theory and modeling; computer simulation
87.19.Hh, % Cardiac dynamics
87.19.Nn  % Electrophysiology
}

\narrowtext

Understanding the dynamics of excitable media such as heart
tissue is a problem of substantial interest to physicists,
physiologists, biomedical engineers, and doctors. For
reasons not yet understood experimentally, the healthy
time-periodic spatially-coherent beating of a human heart
will sometimes change to a nonperiodic spatially-incoherent
fibrillating state in which the heart cannot pump blood
effectively (leading to death if suitable treatment is not
administered quickly). It would be valuable to understand
how the onset of arrhythmias that lead to fibrillation depends on
details such as the heart's size~\cite{Winfree94}, geometry,
electrical state, anisotropic fiber
structure~\cite{Fenton98}, and inhomogeneities. A deeper
understanding of the heart's dynamics may also make possible
the invention of protocols by which electrical feedback
could be used to prevent fibrillation~\cite{control-refs}.

Because of many experimental difficulties in studying the
three-dimensional dynamics of a heart~\cite{exptl-refs},
simulations of cardiac tissue (and more generally of
excitable media) play an extremely important role in
identifying and testing specific mechanisms of
arrhythmia. However, quantitatively accurate simulations of
an entire three-dimensional human heart are not yet
feasible. The essential difficulty is that human heart
muscle is a strongly excitable medium whose electrical
dynamics involve rapidly varying, highly localized fronts
(see Figs.~\protect\ref{fig:1d-front} and~\ref{fig:spirals}). The
width of such a front is about 0.05~cm and a simulation that
approximates well the dynamics of such a front requires a
spatial resolution at least 5~times smaller, $\Delta{x}
\approx 0.01\,\rm cm$. The muscle in an adult human heart
has a volume of about $250 \, \rm cm^3$ and so a uniform
spatial resolution of 0.01~cm would require a computational
grid with $\approx 3 \times 10^8$ nodes. Depending on the
assumed material properties of the heart and on the
quantities of interest to analyze, up to 50~floating point
numbers might be associated with each node, requiring the
storage and processing of about $10^{10}$ numbers per time
step.  The fastest time scale in heart dynamics is
associated with the rapid depolarization of the cell
membrane, about 0.1~msec in duration, and a reasonable
resolution of this depolarization requires a time step about
a fifth of this, $\Delta{t}\approx 0.02\, \rm
msec$. Since arrhythmias such as fibrillation may require
several seconds to become established, the $10^{10}$ numbers
associated with the spatial mesh would have to be evolved
over about $10^6$ time steps.  Such a uniform mesh
calculation currently exceeds existing computational
resources and has not yet been carried out.

A clue about how to improve heart simulations comes from
experiments~\cite{exptl-refs} and
simulations~\cite{Courtemanche96,Fenton98} which suggest
that the electrical membrane potential~$V(t,{\bf x})$ in the
fibrillating state consists of many spirals (for
approximately two-dimensional tissue such as the atrium, see
Fig.~\protect\ref{fig:spirals}) or of many scroll waves 
(for thicker cardiac tissue such as the left
ventricle~\cite{Fenton98}). A striking feature of these
spatiotemporal disordered states is that the dynamics is
{\em sparse}: at any given time, only a small volume
fraction of the excitable medium is occupied by the fronts,
and away from the fronts the dynamics is slowly varying in
space and time. It may then be the case that the
computational effort and storage can be greatly reduced,
from being proportional to the volume of the excitable
medium (the case for a spatially uniform mesh) to being
proportional to the arclength (in~2d) or surface area
(in~3d) of the fronts.

In this Letter, we show for representative solutions of the
quantitatively accurate Luo-Rudy~1 (LR1) membrane model of
cardiac tissue~\cite{Luo91}) that an explicit space-time
adaptive-mesh-refinement algorithm (AMRA)~\cite{Berger-refs}
can indeed take advantage of the sparse excitable dynamics
to reduce by an order of magnitude the computational effort
and memory needed to simulate arrhythmias in large
domains. Further, we show that there is no significant
reduction in accuracy when using an AMRA compared to an
algorithm that uses a spatially uniform mesh at the finest
resolution of the AMRA. Since the AMRA is explicit in time
and has a fairly simple data structure consisting of nested
patches of uniform Cartesian meshes, the AMRA can be
parallelized straightforwardly~\cite{parallel-amr-refs},
leading to a further reduction in computational effort by
the number of processors. The AMRA is also general and does
not require the use of reduced
models~\cite{Courtemanche96,Fenton98}, which increase
efficiency but sacrifice experimental accuracy by using
fewer variables and perhaps explicitly eliminating rapid
variables. The results presented below suggest that a
quantitatively accurate AMRA simulation of fibrillation in
an entire human left ventricle for several seconds with an
effective $0.01 \,\rm cm$ resolution should already be
practical with existing computers. This is highly
encouraging since further improvements to such algorithms
are possible as discussed below.

In the following, we discuss some details of the AMRA and
then its accuracy and efficiency for simulations of the
LR1~model in large one- and two-dimensional domains.  Our
particular algorithm was a straightforward modification of
an AMRA that has been used by other researchers to integrate
hyperbolic sets of partial differential equations such as
the Euler equations of fluid dynamics~\cite{Berger-refs}.
Since key mathematical and algorithmic details are available
elsewhere~\cite{Berger-refs}, only some essential
ingredients and our modifications of them are briefly
described here; a more detailed discussion will be given
elsewhere~\cite{Cherry00}.

The AMRA approximates a given continuous field such as the
cardiac membrane potential~$V(t,{\bf x})$ on a set of nested
locally-uniform patches of $d$-dimensional Cartesian meshes
in a $d$-dimensional Cartesian box~\cite{Berger-refs}. On
each patch, spatial derivatives in the dynamical equations
are approximated by second-order-accurate finite differences
and an explicit method (we use
forward-Euler\cite{FE-comment}) is used to advance in
time. The power of the algorithm arises from its ability to
automatically and efficiently refine or coarsen the
representations of fields by varying the number of grid
points locally to achieve a specified truncation error. A
further reduction in computational effort is achieved by
allowing the the time step to change locally with the
spatial mesh~\cite{Berger-refs}. In related prior work,
Quan et. al.~\cite{Quan98} have studied cardiac models using
spatially adaptive time steps but with a uniform spatial
mesh and alternation of implicit and explicit time steps, 
while Moore~\cite{Moore99fem} has studied reaction-diffusion
equations using a spatially-adaptive fully-implicit method
but with a spatially-uniform adaptive time step. To our
knowledge, ours is the first study of an algorithm for
excitable media for which the spatial and temporal
resolutions change locally.

An important subtlety is that our AMRA was designed for
hyperbolic equations but is here applied to an excitable
medium which is described by {\em parabolic} equations. For
explicit time integrations of hyperbolic equations, the
Courant-Friedrichs-Lewy (CFL) condition for the onset of
numerical instability~\cite{Berger-refs} bounds the largest
possible local time step~$\Delta{t}$ by the first power
of the local spatial resolution~$\Delta{x}$. 
For parabolic equations, the stability condition for an 
explicit algorithm bounds the time step by 
$\Delta{x}^2$ for $\Delta{x}$ sufficiently small.
In the LR1 model, this dependence is evident only for spatial 
resolutions an order of magnitude finer ($\Delta{x} < .0025~cm$) 
than those used in our 
calculations. For resolutions in our range of interest, 
the fast reaction kinetics, not the diffusion operator, 
sets the stability limit on the time step~\cite{Cherry00}.
A standard way to avoid the stability
restriction on~$\Delta{t}$ is to use a semi- or
fully-implicit time-integration
algorithm~\cite{Fenton98,Quan98,Moore99fem}. 
We have estimated 
that by using an explicit integration scheme, 
our time steps on the finest meshes
are about an order of magnitude smaller than those needed to
achieve a 10\% relative error in the speed of the
front (AMRA uses 0.003~ms as opposed to the value 
0.04~ms for the semi-implicit case)~\cite{Cherry00}. 
However, one cannot
conclude that a semi-implicit algorithm is automatically
better than our explicit one since, for a fixed spatial 
resolution, the larger time step allowed by a semi-implicit 
method may give less accuracy during the 
upstroke~\cite{Pollard92} and 
require more computation (some of these issues will be
discussed quantitatively elsewhere for the 1d
case~\cite{Cherry00}). Since the spatiotemporal dynamics of
even the most detailed cardiac membrane models are not yet
understood and the relation between specified local truncation
error and
correct dynamics is also not understood, the present
calculations should be considered as an early but
significant step in finding a good balance between
efficiency and accuracy for simulating arrhythmias in large
domains and over long times.

Our results for the AMRA were obtained for the
quantitatively accurate~LR1 model~\cite{Luo91}, which in~2d
can be written in the form:
%\begin{equation}
%  \begin{eqalign}
%\begin{mathletters}
  \begin{eqnarray}
    C_m \partial_t V(t,x,y) & = &
     {1 \over \beta} \left(
       g_x \partial_x^2 V  +  g_y \partial_y^2 V  \right) 
       -  I_{\rm ion}({\bf m}) - I_{\rm stim}(t,x,y), \nonumber \\
    {d{\bf m} \over dt} & = & {\bf f}({\bf m},V) ,
    \label{eq:lr1-model}
  \end{eqnarray}
%\end{mathletters}
%  \end{eqalign}
%\end{equation}
where $V(t,{\bf x})$ is the membrane potential at time~$t$
and at position~${\bf x} = (x,y)$, $C_m$~is the membrane
capacitance per unit area, $\beta$~is a surface-to-volume
ratio of a heart cell, $g_x$ and~$g_y$~are membrane
conductivities (generally not equal since the heart is
anisotropic), $I_{\rm ion}$~is the total ionic current
flowing across the membrane, and $I_{\rm stim}$ is a specified 
current injected to initiate a propagating wave. 
(For all calculations reported
below, the boundary condition $(\hat{n}\cdot\nabla)V=0$ was
used, where~$\hat{n}$ is the unit vector normal to a given
boundary point.) The seven voltage-sensitive membrane variables
$m_i(t,{\bf x})$ for the LR1~model determine the flow
of various ions across the membrane and satisfy 
{\em ordinary} differential equations, which are also integrated
by a forward-Euler method. The
same membrane parameter values as those of 
Ref.~\protect\cite{Luo91} 
were used except for the calcium conductivity $g_{\rm Ca}$
in the $I_{\rm ion}$ term, whose value was changed from
0.09 to 0.045 (in units of $\rm m\Omega ^{-1}\cdot cm^{-2} $). 
The medium was isotropic with $g_x$ and $g_y$ set to 
1~$\rm { k\Omega ^{-1} \cdot cm ^{-1} }$ and 
$\beta$ set to 3000 $\rm cm^{-1}$.
These values shortened the action
potential duration and led to dynamical states with many
spirals, providing a more challenging test of the AMRA.
% Unless otherwise noted, the medium was
% isotropic with $g_x = g_y$ = 2~k$\Omega^{-1} \cdot $
% cm$^{-1}$ and $\beta$ = 2000 cm$^{-1}$.

In addition to the physical parameters in
Eq.~(\protect\ref{eq:lr1-model}), many numerical and algorithmic
parameters need to be
specified~\cite{Berger-refs,Cherry00}. Several of the more
important choices are an initial resolution for a uniform
coarse mesh covering the domain (we used
$\Delta{x}=0.05\,\rm cm$), the temporal resolution 
for the coarse mesh (we used $\Delta{t}=0.012\, \rm ms$),
the maximum number of grid
levels allowed for refinement (we used the value~3),
the factor by which the spatial mesh is refined locally (we
chose the factor~2), the error tolerance used in the
Richardson extrapolation estimate of the local truncation
error (we chose $\epsilon = 2 \times 10^{-3}$);
and the number of time steps to elapse before
estimating a local error and regridding (we chose 2).

As a first demonstration of the effectiveness of the AMRA,
Fig.~\protect\ref{fig:1d-front} summarizes a 3-level 
calculation of the LR1~model in a 1d~domain of 
length~$L=9\,\rm cm$. The system was stimulated at 
$t=0$ with a $0.2\,\rm cm$ square pulse 
% of height 900 mV
along the left edge of the domain, 
which evolved into a front propagating to the
right (the spatial profile is independent of the initial
condition and of the system size for~$L \ge 9 \,\rm
cm$). One can see from the spatial profile in
Fig.~\protect\ref{fig:1d-front}a at time $t=240\,\rm ms$ 
how narrow is the front (region of depolarization) compared 
to the profile's extent and this specifically is what makes
numerical simulation of highly excitable media so difficult.
In the vicinity of the front, Fig.~\protect\ref{fig:1d-front}b 
shows the grid structure which was automatically calculated by
the ARMA; the colors black, green, and red indicate the coarse,
fine, and finest mesh regions respectively.  Taking into
account the reduction of spatial mesh points and the
asynchronous updating of grid points using spatially varying
time steps~\cite{Berger-refs}, the AMRA overall used a factor 
of~3.6 fewer grid points and did less computational work by 
a factor of 9 
%was overall a factor of~10 more efficient 
for the LR1 model than a
constant-time-step uniform-spatial-mesh forward-Euler code
using the finest space-time resolutions of the AMRA. The
spatial adaptivity of the time step accounts for a factor of~2
in this factor of~10 and so is an important part of the
algorithm. The temporal profiles at a fixed point in space, the 
front speeds, and the times between peak and recovery at a fixed 
point in space (action potential duration) 
for the AMRA and for a high-resolution uniform-mesh
code (discussed in Ref.~\cite{Cherry00}) agree within 0.1\%
relative errors except at the peaks of the temporal profiles, 
where the relative error is about 4\%. We conclude 
that there is no significant loss of
accuracy when using the more efficient AMRA.

Fig.~\protect\ref{fig:spirals} shows how the AMRA performs 
for the LR1~model in a large square domain of size~$L=8\,
\rm cm$, using the same parameter values as the 1d case, 
for which spirals are unstable and break up
into other spirals. This complex many-spiral dynamical
state
%~\cite{Strain98prl} 
is a much stronger test of the
efficiency and utility of an AMRA than 
Fig.~\protect\ref{fig:1d-front}
since the geometry of the fronts fluctuates strongly in
time.  A multi-spiral state was initiated by a standard
S1-S2 stimulation protocol~\cite{Courtemanche96} in which a
right-going planar pulse is created by stimulating the 
left edge of the domain (the S1~stimulus), and 
the lower left quadrant of the domain is excited (the S2~stimulus)
334~ms later, when the left half of the domain has returned to 
rest but the right half is still repolarizing. 
% The initial coarse uniform spatial mesh had
% resolution~$\Delta{x} = 0.05\,\rm cm$ and the initial
% uniform time step was~$\Delta{t}=0.012\,\rm ms$ and these
% values were resolved down to $\Delta{x}=0.0125\,\rm cm$
% and~$\Delta{t} = 0.003\,\rm ms$ at the finest local
% levels. The simulation lasted 1800~ms.
A comparison of the field~$V$ with the instantaneous grid
structure approximating~$V$ is given in 
Fig.~\protect\ref{fig:spirals} 1346~ms after S2 
and demonstrates how the 
AMRA is able to increase automatically the space-time resolution
only in the vicinity of the fronts, greatly decreasing the
overall computational effort since, at any given time, the
sharp fronts indeed occupy only a small fraction of the
domain. The total
number of mesh points used by the AMR varies substantially
with time, from $3 \times 10^4$ to $7 \times 10^4$ mesh 
points with an average of $5 \times 10^4$.  A
comparison of these results with those required by a
uniform-spatial-mesh constant-time-step code using the
finest AMRA resolution~\cite{Cherry00} shows that the AMRA
uses about 8~times fewer mesh points, requires less 
integration work by a factor of 12, and achieves a speedup
of about a factor of~11~\cite{Cherry00}.

The above results can be used to estimate the computer
time needed by the ARMA to integrate for one second the LR1 
model for a 3d~section of left ventricular 
wall of dimensions $8 \, \rm cm \times 8 \, \rm cm
\times 1 \, \rm cm$, with an effective fine
uniform mesh resolution of~$\Delta{x}=0.0125 \, \rm cm$
in space and $\Delta{t}=0.003\,\rm msec$ in time. 
On a Pentium~III 500~MHz computer, we found that a 3-level 
2d~AMRA calculation at this resolution takes about 3~days. 
The time for the 3d~calculation then can be estimated by assuming 
that each of the spirals in Fig.~\protect\ref{fig:spirals} becomes
a continuous stack of spirals (a scroll wave), with the 
stack transverse to the square sides of the domain~\cite{Fenton98},
and correspondingly that the mesh refinements extend uniformly
from the 2d~case through the transverse direction. A 3d~AMRA 
calculation should then take roughly 
15~days, which is a factor of~17 speedup
over the 9~months required to complete a similar calculation 
using a uniform space-time mesh with the above resolution.
Without substantial change to the AMRA, an additional speedup
of at least~10 can be gained by using a distributed parallel 
computer with 100~Pentium~III processors, and another speedup 
of~5 by using table-lookups to avoid the many exponentiations
associated with the integration of the membrane 
variables~$m_i(t)$.
These further gains would reduce the total simulation time 
for one second of the LR1~model in this 3d~domain to 7~hours 
or less. (With a substantial modification to make the AMRA 
semi-implicit, another reduction by a factor of 2-3 might be 
possible.)
Simulation of an entire heart (a factor of 4~greater in volume)
for one second with a LR~1 model should then be possible on 
the time scale of one day, which is acceptably fast for exploring 
many interesting questions about the dependence of arrhythmias 
on parameters.

In summary, we have shown that an explicit space-time
adaptive algorithm~\cite{Berger-refs} using one of the
simplest possible data structures (a hierarchy of Cartesian
meshes) can already attain an order of magnitude reduction
in computational effort and memory when applied to the
experimentally based LR1 cardiac membrane
model~\cite{Luo91}, and that this reduction is achieved
without incurring a corresponding reduction in accuracy when
compared to an explicit code using a uniform space-time
mesh. 
%Important next questions to investigate are whether
%the algorithm can be improved by using implicit time
%integration, how to generalize the method to curved
%boundaries, and making specific applications
%to the initiation and control of human arrhythmias.
Important next steps include determining whether the 
algorithm can be improved by using implicit time integration, 
generalizing the method to curved boundaries, and making 
specific applications to the initiation and control of human 
arrhythmias.

We thank M.~Berger, Z.~Qu, and A.~Garfinkel for useful
discussions and especially M.~Berger for making available to
us one of her AMRA codes. This work was supported by a NSF
Graduate Research Fellowship, by NSF grant DMS-9722814, and
by NIH grant R29-HL-57478.

% For final version, comment out \bibliography lines and
% uncomment the \input{references} line.

%\bibliographystyle{prsty} % entries in order of citation
%\bibliography{cardiology,cfd,control,excitable-media,na,spatiotemporal}

\begin{references}
\bibitem{FE-comment} For strongly excitable media with fast
time scales, the simple first-order-accurate forward-Euler
method is acceptable since the fronts evolve so quickly that
a time step small enough to resolve these fronts also leads
to good accuracy.
\end{references}
%\input{references}

\newpage

\begin{figure}   % fig 1
\caption{ {\bf (a)} Spatial profile $V(t,x)$ at time 
$t=240\,\rm
ms$ for a 1d~front propagating to the right in a
domain of length~$L=9 \, \rm cm$, as calculated by a
3-level adaptive mesh refinement algorithm (AMRA) for the
Luo-Rudy~1 (LR1) cardiac model~\protect\cite{Luo91}.
% Stimulus was applied over the left 0.2 cm of the domain
% for a duration of 0.1 ms.
The three regions of coarse, fine, and finest mesh
resolution (from $\Delta{x}=0.05\,\rm cm$,
$\Delta{t}=0.012\,\rm ms$ to $\Delta{x}=0.0125\,\rm
cm$, $\Delta{t}=0.003\,\rm ms$) are indicated by the
black, green, and red portions of the curve. {\bf (b)}
Blowup of the small interval indicated near $x=8.4\, \rm cm$ 
in {\bf
(a)}, showing how the 3-level mesh structure (vertical
lines) has automatically resolved the sharp front.
% {\bf (c)} Temporal profile of~$V(t,x)$ at fixed 
% position~$x=8.4\, cm$ as calculated by a 3-level AMR
% code with the same parameters as in {\bf (a)}. {\bf (d)}
% Relative error of the AMR temporal profile~$V(t,x)$ in
% {\bf (c)} when compared with the profile calculated by a
% uniform-mesh code at the finest resolution of the AMR code.
% $\|V_{\rm AMR}(t,x) - V_{\rm uniform}(t,x)\|/105 \, \rm mv$
% between profiles calculates for the 3-level AMR code and for
% uniform-grid code at the finest AMR level.
% relative error in velocity < 0.1%
% relative error in APD < 0.01%
% relative error in action potential shape < 0.1% except
%  at the peak where relative error is about 4%.
}
\label{fig:1d-front}
\end{figure}

\begin{figure}   % fig2
\caption{ {\bf (a)}: Three-level AMRA calculation of the 2d
LR1~model at time $t=1346\,\rm ms$ after stimulus~S2, in a
square domain of length~$L=8\,\rm cm$.  Field value ranges
for~$V(t,x,y)$ are color coded with blue for $V \ge -5\,\rm
mV$, red for $-5 \le V \le -65\,\rm mV$, and yellow for $V
\le -65\,\rm mV$.  Parameter values are the same as in
Fig.~\protect\ref{fig:1d-front}. {\bf (b)}: The hierarchical Cartesian
meshes of the AMR algorithm corresponding to the snapshot
of~$V$ in {\bf (a)}. The yellow and green regions correspond
to the fine (level~2) and finest (level~3) grids and track
closely the fronts. }
\label{fig:spirals}
\end{figure}

%\begin{figure}  % fig 3
%\caption{ Number of spatial mesh points used by the AMRA as
%a function of time for the 3-level calculation described in
%\figref{fig:spirals}.  }
%\label{fig:mesh-points-vs-time}
%\end{figure}

%\newpage
%\centering\includegraphics[width=5in]{figures/fig1.ps}
%\begin{center}
%\large\bf Figure 1
%\end{center}
%
%\newpage
%\vspace{.5in}
%\centering\includegraphics[width=4in]{figures/fig2a.ps}
%\centering\includegraphics[width=4in]{figures/fig2b.ps}
%\begin{center}
%\large\bf Figure 2
%\end{center}
%
%\newpage
%\centering\includegraphics[width=4in]{figures/fig3.ps}
%\begin{center}
%\large\bf Figure 3
%\end{center}

\end{document}